\input{aipcheck}

\documentclass[
    ,final            
  ]
  {aipproc}

\layoutstyle{8x11double}
\usepackage{unites2e}

\begin{document}

\title{Status of the VERITAS Observatory}


\author{J.~Holder}{
  address={Dept. of Physics and Astronomy and the Bartol Research Institute, University of Delaware, DE 19716, USA}
}

\author{V.A.~Acciari, E.~Aliu, T.~Arlen, M.~Beilicke, W.~Benbow, S.M.~Bradbury, J.H.~Buckley, V.~Bugaev, Y.~Butt, K.L.~Byrum, A.~Cannon, O.~Celik, A.~Cesarini, L.~Ciupik,
Y.C.K.~Chow, P.~Cogan, P.~Colin, W.~Cui, M.K.~Daniel, T.~Ergin,
A.D.~Falcone, S.J.~Fegan, J.P.~Finley, G.~Finnegan, P.~Fortin, L.F.~Fortson,
A.~Furniss, G.H.~Gillanders, J.~Grube, R.~Guenette, G.~Gyuk, D.~Hanna,
E.~Hays, D.~Horan, C.M.~Hui, T.B.~Humensky, A.~Imran,
P.~Kaaret, N.~Karlsson, M.~Kertzman, D.B.~Kieda, J.~Kildea,
A.~Konopelko, H.~Krawczynski, F.~Krennrich, M.J.~Lang, S.~LeBohec,
G.~Maier, A.~McCann, M.~McCutcheon, P.~Moriarty, R.~Mukherjee,
T.~Nagai, J.~Niemiec, R.A.~Ong, D.~Pandel, J.S.~Perkins, M.~Pohl,
J.~Quinn, K.~Ragan, L.C.~Reyes, P.T.~Reynolds, H.J.~Rose,
M.~Schroedter, G.H.~Sembroski, A.W.~Smith, D.~Steele, S.P.~Swordy,
J.A.~Toner, L.~Valcarcel, V.V.~Vassiliev, R.~Wagner, S.P.~Wakely,
J.E.~Ward, T.C.~Weekes, A.~Weinstein, R.J.~White, D.A.~Williams,
S.A.~Wissel, M.~Wood, B.~Zitzer }
{ address={see \texttt{http://veritas.sao.arizona.edu} for a full list of
affiliations} }

\begin{abstract}
VERITAS, an Imaging Atmospheric Cherenkov Telescope (IACT) system for
gammma-ray astronomy in the GeV-TeV range, has recently completed its
first season of observations with a full array of four telescopes. A
number of astrophysical gamma-ray sources have been detected, both
galactic and extragalactic, including sources previously unknown at
TeV energies. We describe the status of the array and some highlight
results, and assess the technical performance, sensitivity and shower
reconstruction capabilities.
\end{abstract}

\maketitle


\section{Introduction}

The current generation of TeV gamma-ray telescopes, H.E.S.S., MAGIC
and VERITAS, provide an order of magnitude sensitivity improvement
over previous instruments, and have dramatically increased the number
of astrophysical gamma-ray sources, and source classes, available for
study. VERITAS is the most recent of these observatories to come
online, and has now completed its first full season of observations.

\begin{figure}
  \includegraphics[width=1.0\textwidth]{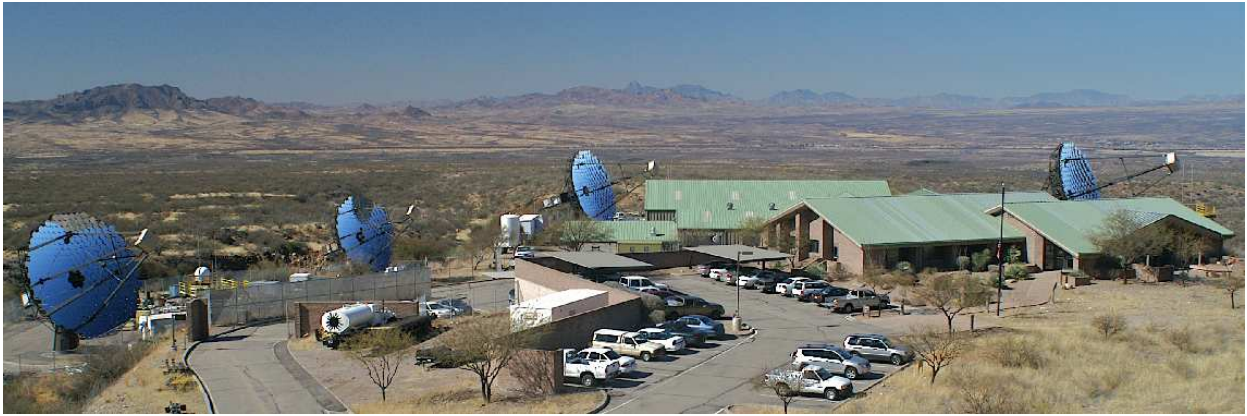}
  \caption{   
  \label{array}
  The VERITAS Array}
\end{figure}

The VERITAS array \cite{Holder06} comprises four, $12\U{m}$ diameter
imaging atmospheric Cherenkov telescopes located in Tucson, Arizona
($31^{\circ}40'30''\U{N}, 110^{\circ}57'07''\U{W}$, $1268\U{m}$ above
sea level) (Figure~\ref{array}). Each telescope is equipped with a
pixellated photomultiplier tube (PMT) camera, providing a
$\sim3.5^{\circ}$ field of view.  The array layout gives telescope
baselines varying from $35\U{m}$ to $108\U{m}$. Construction of the
array began with a prototype instrument in Fall 2003; the first full
telescope was commissioned in winter 2005, and the full array
completed in spring 2007.

VERITAS observations are made under clear sky conditions from
September to June, with a summer shutdown due to local
monsoon conditions. The 2007-2008 observing season provided
$\sim700\U{hours}$ of good, dark sky observations as well as
$\sim100\U{hours}$ taken under partial moonlight. Moonlight
observations are generally made with the standard high voltage and
trigger conditions, and provide a valuable increase in duty cycle:
observations of the newly discovered TeV AGN WComae \cite{Acciari08a},
made under moonlight, led to the detection of a flare with a flux of
$\sim25\%$ of the Crab Nebula flux \cite{WComaeAtel}.

The Whipple $10\U{m}$ telescope operates on a similar schedule at a
distance of $7\U{km}$ from the VERITAS site, and provides a trigger
for high flux ($>10\%$ Crab flux) transient events in known TeV
sources. A recent Whipple $10\U{m}$ trigger led to the prompt
measurement by VERITAS of an extreme flaring state in the TeV blazar
Markarian 421, peaking at $\sim10\U{Crab}$ \cite{Reyes08}.

\section{Performance}
Each PMT pixel in the telescope cameras is instrumented with a
custom-built 500 Mega-samples per second Flash ADC
\cite{Buckley03}. The FADC traces are read-out when the array trigger
conditions are met (two out of four telescopes triggered, each with
the signals from at least three adjacent PMTs crossing their
individual constant fraction discriminator thresholds). In the first stage of the
analysis, the signal traces are integrated. The resultant shower
images are then calibrated, cleaned and parameterized. Gamma-ray candidate events
are selected based primarily on the image shape (using ``reduced
mean-scaled \textit{width} and
\textit{length}'' parameters) and the reconstructed angular distance
from the candidate source location, $\theta$ (\cite{Acciari08b}, and
references therein). The following sections present some of the
critical performance metrics of the system.
\vspace{-0.3cm}
\subsubsection{Angular Reconstruction}
\begin{figure}
  \includegraphics[width=0.5\textwidth]{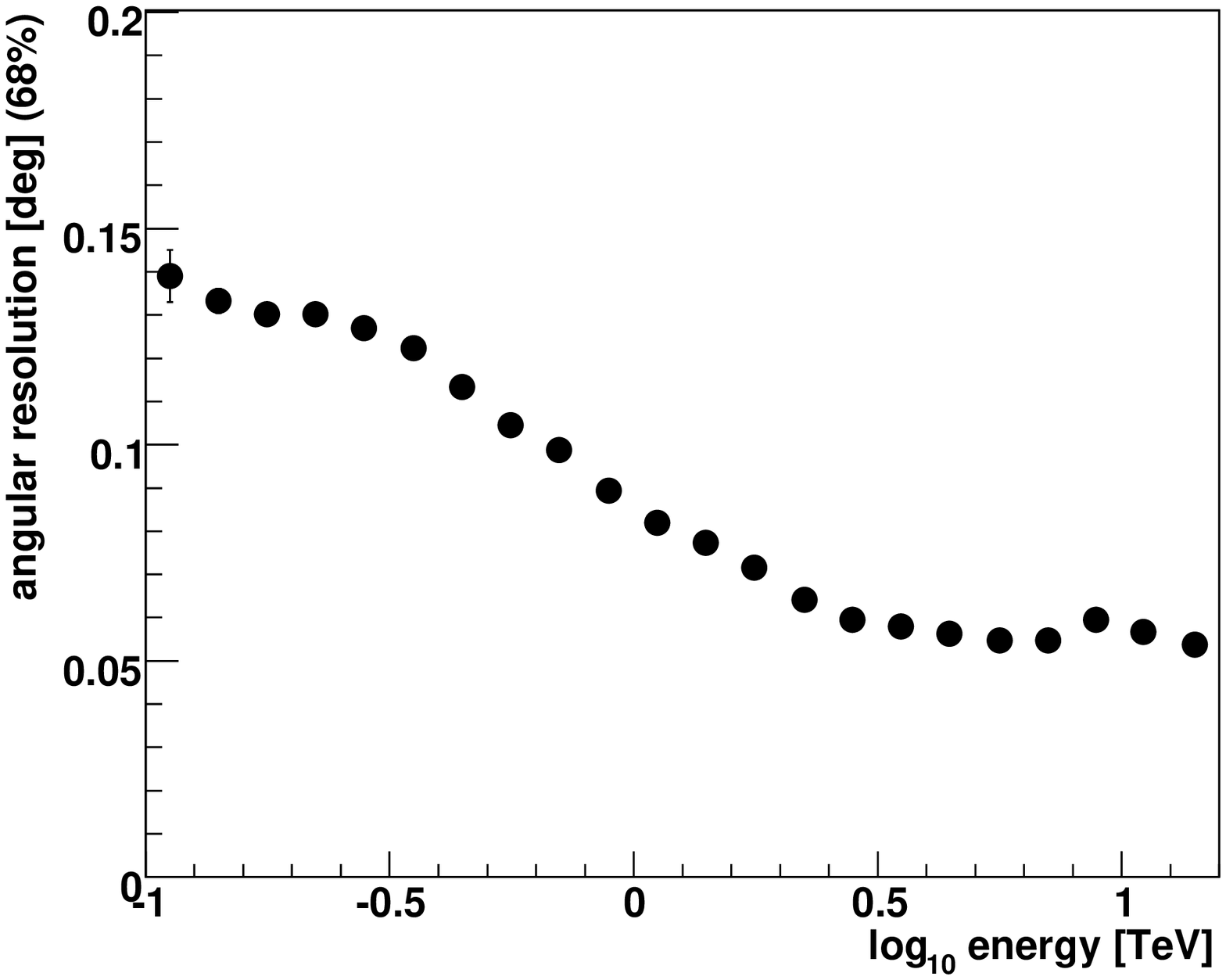}
  \includegraphics[width=0.5\textwidth]{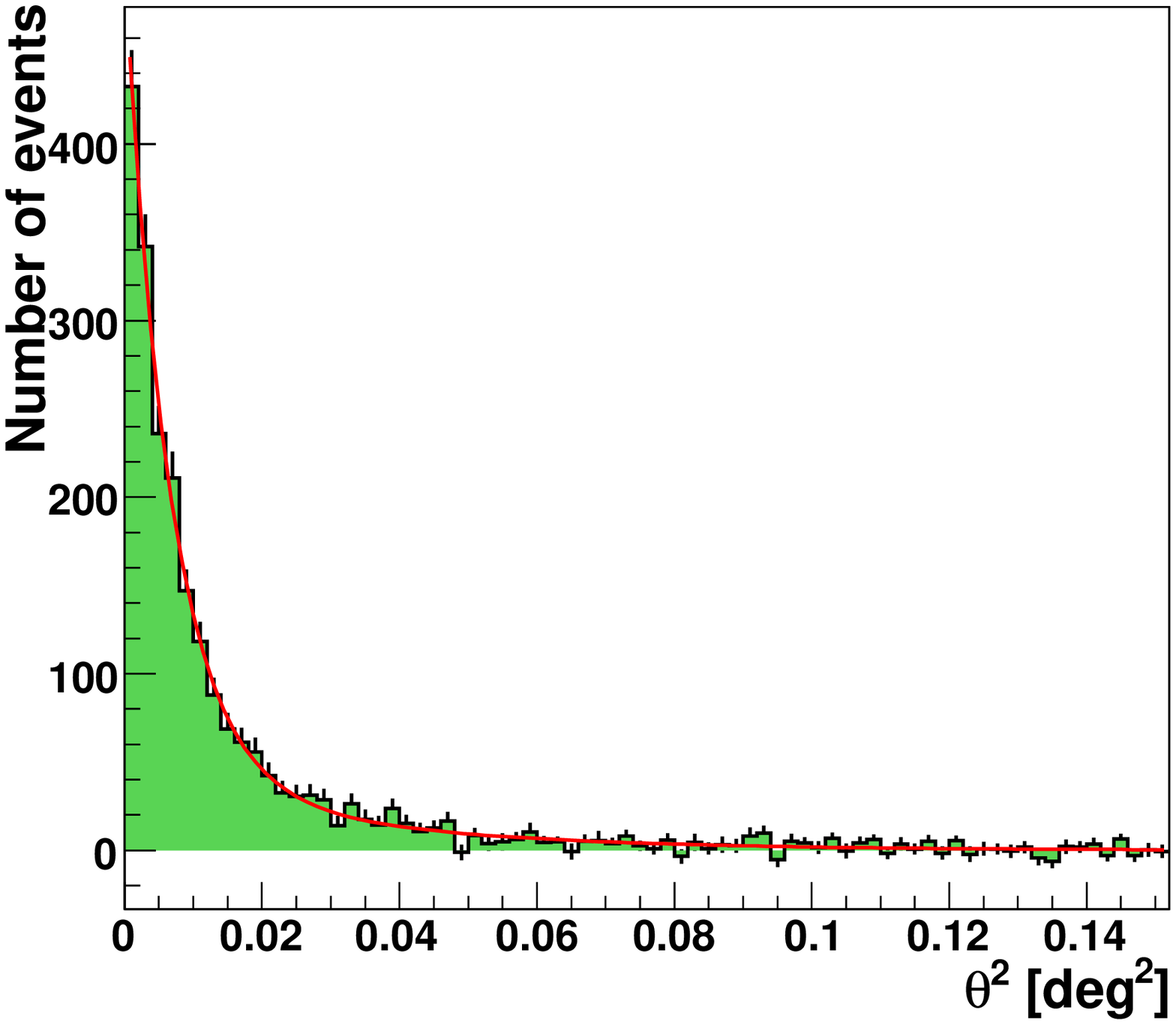}
  \caption{ \label{angres} 
{\bf Left:} The angular resolution as a function of energy, measured using simulated gamma-ray events.
{\bf Right:} The $\theta^2$ distribution for the gamma-ray excess measured during observations of the Crab Nebula. The fit is a combination of two Gaussian functions, as described in \cite{Lemoine06}  
}
\end{figure}
Angular resolution is a function of both the analysis procedure and
the absolute pointing accuracy of the telescopes in the
array. Calibration of the telescope pointing is performed on a monthly
basis, and the results input to a pointing model in the telescope
tracking software which corrects for the measured misalignments and
flexures during observations. This provides a systematic pointing
accuracy of $\sim0.02^{\circ}$. Recently installed optical pointing monitor
telescopes provide further improvement offline.

The optimum angular resolution and sensitivity to weak sources is
achieved when relatively strict gamma-ray selection cuts are
applied. The results shown here require that all four telescopes are
involved in the shower reconstruction, and that each image has an
integrated charge (\textit{size}) per telescope of $>75$
photo-electrons.  With these requirements we obtain the energy
dependent angular resolution illustrated in
Figure~\ref{angres}. Averaged over all energies, the angular
resolution for a single event is $\sim0.1^\circ$ (68\% containment radius). This source
localization capability allows us to distinguish between point sources
and sources with a small angular extent, as illustrated by VERITAS
observations of the extended TeV source in the IC443 supernova remnant
\cite{Humensky08}.

\begin{figure}
  \includegraphics[width=0.5\textwidth]{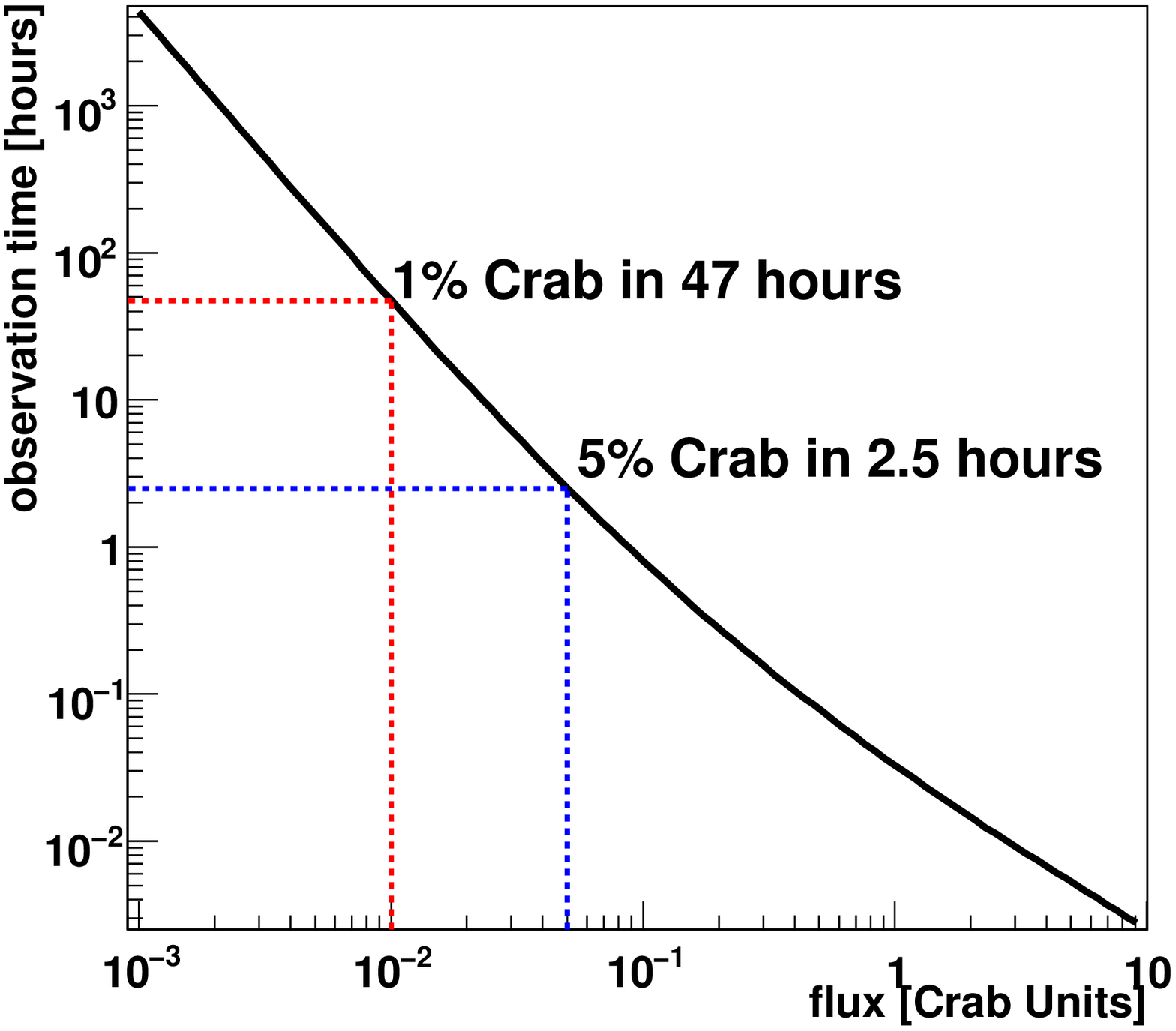}
  \includegraphics[width=0.5\textwidth]{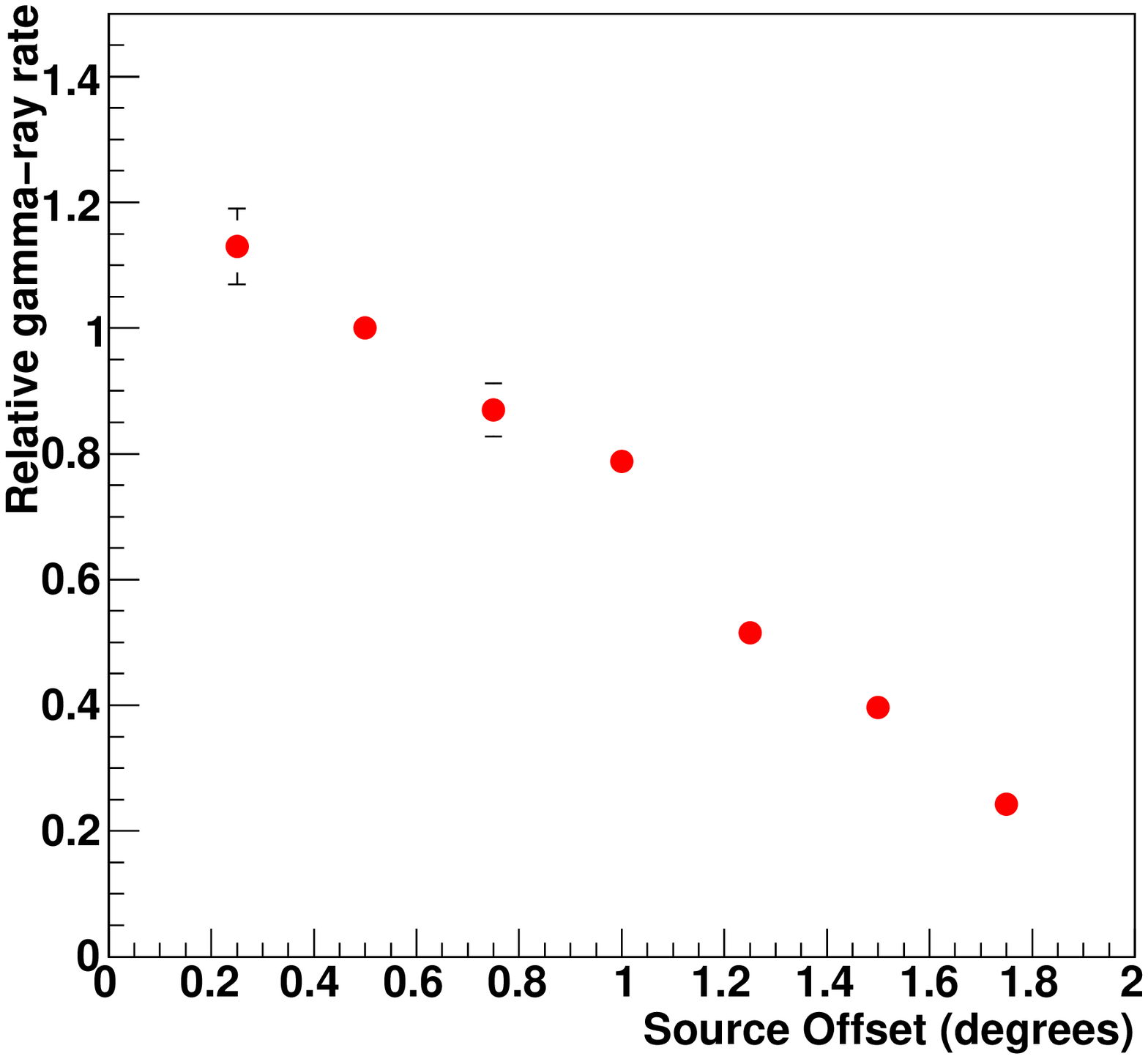}
  \caption{ \label{sensitivity} 
{\bf Left:} The time required for a 5$\sigma$ detection for different source strengths, assuming a Crab-like gamma-ray spectrum.
{\bf Right:} The gamma-ray rate as a function of source position, measured using observations of Crab Nebula. The measurements are normalized to the measured rate at an offset of $0.5^{\circ}$.
}
\end{figure}

\vspace{-0.3cm}
\subsubsection{Sensitivity}
Using the same strict cuts, with the addition of an angular cut of
$\theta<0.12^{\circ}$, Figure~\ref{sensitivity} illustrates the
current sensitivity of the array as measured using Crab Nebula
observations. An unresolved source with a flux of 1\% that of the Crab
Nebula is detected in 47 hours of observations, a 5\% Crab source
requires 2.5 hours. The TeV blazar 1ES~0806+524 has been detected by
VERITAS at this flux level
\cite{Cogan08}. Figure~\ref{sensitivity} also shows the gamma-ray rate as
a function of the source position in the field of view. The array has
useful sensitivity out to at least $2.0^{\circ}$, which has allowed
the measurement of a gamma-ray flux from 1ES~1218+304 during
observations centered around WComae, separated by $1.94^{\circ}$
\cite{Acciari08a}

\begin{figure}
  \includegraphics[width=0.8\textwidth]{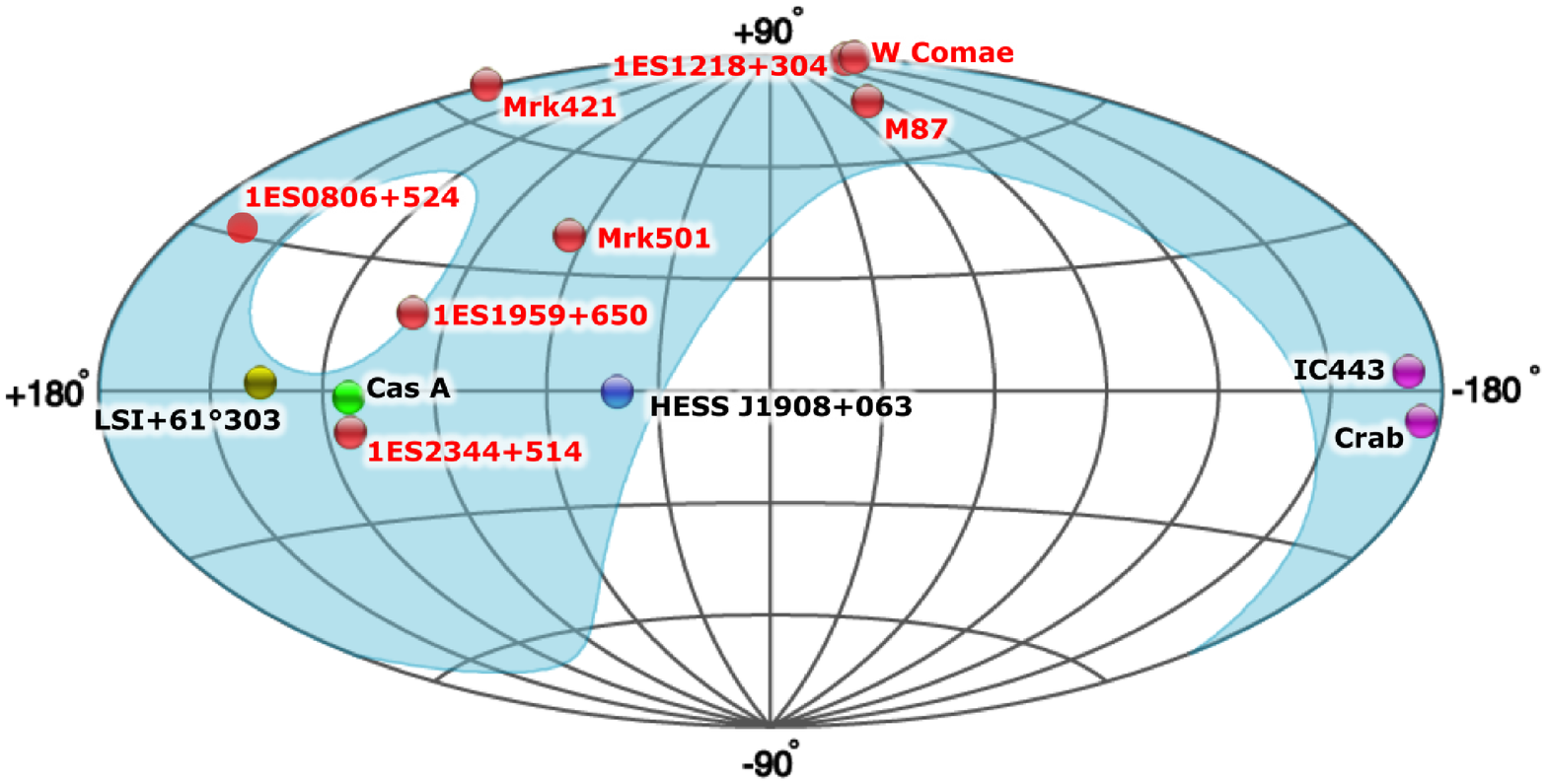}
  \caption{ \label{veritas_sky} The VERITAS source catalogue as of
  summer 2008 (in galactic coordinates). Extragalactic sources are
  labelled in red, those within our galaxy are labelled in black. The
  shaded area shows the region of the sky visible to VERITAS at
  elevations greater than $55^{\circ}$. Figure provided courtesy of
  TevCat (\texttt{http://tevcat.uchicago.edu/}) }
\end{figure}

\vspace{-0.3cm}
\subsubsection{Energy Reconstruction}
The energy of incident gamma rays is calculated from the measured
impact parameter, image \textit{size}, and azimuth and elevation angle
using lookup tables filled from simulated gamma-ray events. The energy
scale of the simulations is calibrated using a variety of
complementary methods, such as \textit{in situ} measurements of the
PMT single photo-electron response, studies of the light yield from
local muons, and measurements of the Rayleigh scattered light from a
nearby laser pulse \cite{Hui08}. In order to provide a wide dynamic
range for the energy spectra, gamma-ray selection cuts are loosened to
allow any event with at least two telescopes recording an image with
>4 pixels, excluding the telescope combination with the smallest
baseline (Telescopes 1 and 4). Spectral reconstruction is possible
from a minimum energy of $\sim150\U{GeV}$ and with an energy
resolution of $\sim15\%$ at high energies.


\section{Summary}

We have completed the first full season of observations with the
four-telescope VERITAS array. The technical performance, site
conditions and mechanical reliability are good, providing
$\sim800\U{hours}$ of clear sky observations each year. Highlight
results from this first season include the detection of two new TeV
sources, 1ES~0806+524 and WComae, and detailed studies of the
supernova remnants IC443 and Cas A, including a measurement of an
extended TeV source in IC443 \cite{Humensky08}.  Extensive
multiwavelength observations have been made of known extragalactic TeV
sources, including a comprehensive multi-instrument campaign on
M87 \cite{Beilicke08} the detection of extreme flaring from
Markarian~421 and the highest recorded flux from
1ES~2344+514 \cite{Cogan08}. Additional observations include uniquely
interesting galactic sources such as the TeV binary
LS~I$+61^{\circ}303$ \cite{Maier08} and HESS~J1908+063 \cite{Ward08},
identified with the Milagro source MGRO~J1908+06.

We conclude by presenting the VERITAS source catalogue at the close of
this first season (Figure~\ref{veritas_sky}), containing 13 northern
hemisphere TeV sources. Ongoing projects include a survey of the
Cygnus region of the galactic plane, and a search for gamma-ray
signatures from objects predicted to have an enhanced dark matter
component.

\begin{theacknowledgments}
This research is supported by grants from the U.S. Department of
Energy, the U.S. National Science Foundation, the Smithsonian
Institution, by NSERC in Canada, PPARC in the U.K. and by Science
Foundation, Ireland.
\end{theacknowledgments}

\bibliographystyle{aipproc}   


\end{document}